\documentstyle[12pt]{article}
\begin{document}

\title{Isologous Diversification: A Theory of Cell Differentiation }
\author{
        Kunihiko Kaneko\\
        {\small \sl Department of Pure and Applied Sciences}\\
        {\small \sl University of Tokyo, Komaba, Meguro-ku, Tokyo 153, JAPAN}\\
       and\\
        Tetsuya Yomo\\
        {\small \sl Department of Biotechnology}\\
        {\small \sl Faculty of Engineering }\\
        {\small \sl Osaka University 2-1 Suita, Osaka 565, JAPAN}
\\}

\date{}
\maketitle
\begin{abstract}

Isologous diversification theory for cell differentiation is
proposed, based on simulations of interacting cells with
biochemical networks and cell division process following consumption of
some chemicals.
According to the simulations of the interaction-based dynamical systems
model,
the following scenario of the cell differentiation is proposed.
(1) 
Up to some threshold number, divisions bring about almost identical 
cells with synchronized biochemical oscillations.
(2)As the number is increased the oscillations lose the synchrony, leading to
groups of cells with different phases of oscillations.
(3)Amplitudes of oscillation and averaged chemical compositions start to 
differ by groups of cells.
The differentiated behavior of states is transmitted to daughter cells.
(4)Recursivity is formed so that the daughter cells keep the identical 
chemical character.  This ``memory" is made possible through the
transfer of initial conditions.
(5) Successive differentiation proceeds.

Mechanism of tumor cell formation, origin of stem cells, anomalous 
differentiation by transplantations, apoptosis and other features of cell 
differentiation process are also discussed, with some novel predictions.

(Keywords: differentiation, chemical network, cell division,
clustering, open chaos)
\end{abstract}

\section{Introduction}

\subsection{Biological background}

Development of organisms from their fertilized eggs is one of the most
elegant emergence in biology and has been investigated by many cellular and
molecular biologists to elucidate how different types of cells appear and
organize beautiful structure of a matured body.  
Many of the essential genes for the body plan have been identified.
Each gene, responding to the products from the other
genes, turns on or off so as to give
different physiological states and to  produce a variety of  cells.

        The gene network picture is expressed by 
``canalization".  Depending on initial conditions, there are a variety of
final states as is expressed schematically by the landscape in Fig. 1a).  In
the term of dynamical system theory, there are many attractors in the
network system.  If the initial condition for canalization is given by the
initial gene expression and/or by the environments, a cell differentiates
into one of the attractors according to their basins.  A beautiful and 
pioneering study is given by Kauffman (1969) to demonstrate that the 
Boolean networks
of genes have a variety of final states.  His work clearly shows that 
with various initial conditions the  gene  network
leads to the existence of a variety of cell types under a  single
external condition.
However, needless to say, a single initial condition embedded in a fertilized
egg produces several different cell types.  Thus, the following
essential question towards the
gene network picture remains: How do the different initial and/or external 
conditions arise,
leading to different cell types through the developmental process?
How can the differentiation process be robust against perturbations
to initial and external conditions?

There are some experiments which point out the relevance of
cellular interactions to internal states of a cell and
to the robustness against perturbations.
One of the authors and his colleagues reported (Ko et al., 1994)
that even under a
single external condition cells differentiate to some distinct 
physiological states.  In their experiment, {\it E. coli}
was successively cultivated
in a well-stirred liquid culture in order to impose the same external
condition on each of the cells.  The population of the {\it E. coli} was 
shown to
include distinct cell types. The fraction of each  cell type exhibited
a complex oscillation in the time course.  Moreover, it was
shown that in a few repeats of the single colony isolation, 
some colonies inherit the
physiological state, while the others present the state other than that their
parental colonies exhibit.  It is unlikely that each cell of {\it E.coli} in 
the culture from a single cell exhibits different initial conditions in its 
gene network. Thus, the experiments show that under the same initial and
external conditions, the cells can autonomously differentiate.

     It has been established that by transplant experiments, some
cells,  changing their  fate, dedifferentiate and  come  to   a
different  cell  type. Therefore, intercellular  relationship  is
essential  to the determination of a cell type. The  experimental
results  show  that  the differentiation process  is  dynamic  in
principle,  and  the  fate of a cell  is  determined  dynamically
through  the interactions with environment or with  other  cells.
Rubin, in a series of papers, has shown that a cell line(  NIH
3T3)  from a mouse epigenetically transforms to some  different
types of foci in size under the same condition (Yao and Rubin 1994;
Chow et al., 1994, Rubin 1994a,1994b).  
In addition, the frequency of transformation and the type of transformed cells
depend on the cell density and the history of the cell culture
before the transformation.  This, of course, does not deny 
possible roles of mutations but it at least suggests the relevance of 
intercellular interactions to the cell transformation or differentiation.
The importance of interaction to differentiation has also been pointed out,
for example, by Goodwin (1982), and Newman (1990).
However we believe that
none  has  taken fully into account of interaction-induced  and  dynamic
viewpoints successfully.  We, hereby  propose a novel theory of cell 
differentiation  
based  on a dynamic and  cell-interaction viewpoint.

It may be useful to state the standpoint of our theory and
modeling, in advance.  We do not aim at giving a model
with one-to-one correspondence with biological facts at the present.
Rather we present an abstract model and discuss its general features,
to show that a prototype of cell differentiation emerges
even without implementing a programmed switching process of
genes.  It should be noted that
the differentiation progresses with autonomous choice
of initial conditions of internal cellular states through interactions.
The process
is shown to be a general consequence of dynamical systems of
reproducing and interacting units
with internal degrees of freedom.

To make one-to-one correspondence with experimental observations,
models with detailed physiological information must be required.
Our model may look rather premature with regards to
such correspondence.  Hoewever, on the other hand, our theory and modeling 
are essential to understand how a cell society and differentiated 
multi-cellular organisms
emerge without sophisticated programs implemented in advance.
The theory also provides a novel viewpoint to several problems in cell biology,
such as the transformation and apoptosis.

\subsection{Isologous diversification theory}

In the present paper we propose a novel viewpoint of
cell differentiation, which satisfies the dynamic, and interaction-based
picture and also explains the inheritance of cell types through an initial
condition of chemicals in cells.

The background of this theory lies in the dynamic clustering
in globally coupled chaotic systems (Kaneko,1989,90,91,92), where
chaos leads to differentiation of identical elements
through interaction among them.
The relevance of dynamic change of relationships among elements
to biological networks has been discussed (Kaneko 1994).
Even if each oscillatory element does not show chaotic behavior,
the dynamic clustering appears when phase differences of oscillators are 
amplified through the interaction among them.
Cell differentiation provides such an example (Kaneko and Yomo 1994).

One important missing factor in the dynamical systems theory is
the change of degrees of freedom.  Cells divide to create
a new set of dynamical variables.  Previously we introduced the
term ``open chaos" to address the instability 
in conjunction with the
change of the degrees of freedom (Kaneko and Yomo 1994).
In open chaos small deviation is amplified, which 
finally leads to the change in the
dimension of the phase space itself  unlike those in chaos.

Based on these dynamical systems studies and simulations of the
cell differentiation model ( to be presented),
we propose ``isologous diversification theory" as a general
mechanism of spontaneous differentiation of replicating
biological units.  Here we adopt the term ``isologous", in 
contrast with ``homologous", to stress
our general mechanism that any  ``identical" (rather than similar)
units naturally differentiate through interactions \footnotemark.  
It is useful to describe the basic framework of the
theory here to facilitate the understanding of the later simulation results. 
The ``isologous diversification" is summarized as follows
(see also Fig.2 for schematic representation):

\footnotetext{ The use of the term ``isologous" is suggested by
Susumu Ohno.}

\vspace{.2in}
{\sl Let us take biological units ( e.g. cells),
interacting with each other, and with ability of reproduction.  
A state of each unit has internal dynamics (e.g., biochemical reaction) 
which allows for 
nonlinear oscillation ( through e.g., autocatalytic reaction).
Through this inter- and intra- unit dynamics,
the total system consists of coupled nonlinear oscillator units.
As the number of
units increases by the reproduction, they are differentiated
spontaneously, through the following stages\footnotemark.

{\bf (1)  Synchonous oscillations  of  identical units.}
Up to some threshold number of units, all of them oscillate synchronously,
and their states are identical.  

{\bf (2)  Differentiation  of the phases of oscillations  of  internal
states.}
When the number of units exceeds the threshold, they lose identical and
coherent dynamics.
Although the state of each unit is different at an instance, 
averaged behaviors over oscillation periods are essentially the same.
Only the phase of oscillations differs by units.

The emergence of the stage is a general consequence of
the dynamic clustering (Kaneko, 1989,90). 
It is expected that the oscillators split into clusters
with different phases of oscillations, when there
is strong interaction among them.  

{\bf (3)  Differentiation  of the amplitudes of internal  states. }
At this stage, the states are different even after taking the temporal average
over periods.
It follows that the behavior of states
(e.g.,  composition of chemicals, cycles of oscillations, and  so
on) is differentiated.  The clustering of units with regards to amplitudes 
here is again a nature of coupled oscillators when the
interaction is suitably chosen.

{\bf (4)  Transfer  of the differentiated state to the  offsprings  by
reproduction. }

Each type of a differentiated cell is preserved to its offsprings.
Chemical composition of a cell attains recursivity with respect to divisions. 
Thus a kind of ``memory" is formed, through the
transfer of initial conditions (e.g., of chemicals) 
during the reproduction ( e.g., cell division).
By reproduction, the initial condition of a unit is
determined to give a unit of the same  type at the next generation.  

{\bf (5)  Hierarchy of organized groups. }
This stage is the  result  of
successive differentiation with time.  Thus, the total system consists of 
units (cells) of diverse behaviors, leading to a heterogeneous society. }

\footnotetext{The later stages are not necessarily chronologically separated.
The stage (5) often proceeds with (3) and (4), while the stage (4) can
occur with (3).}

\vspace{.2in}

As mentioned, the above stages are
based on the general features of coupled dynamical systems. 
With the reproduction of units, the interaction among them gets stronger
and leads to successive diversification of the behavior.
Thus the identical elements tend to be diversified
through the interplay of nonlinear oscillations,  interaction, and
the change of degrees of freedom (e.g., the number of cells).

The  fourth stage is conceptually new.
The observed memory there lies not solely in the internal dynamics but
also  in the interactions among the units.  If one  is  concerned
with  the internal dynamics, the memory should be  determined  by
the basin for the attractors of internal dynamics, as in Fig. 1a). 
The final destination of the balls in the figure corresponds to
the memory.
However, there remains two factors that need to  be
considered.   One  is the division process, which  increases  the
degrees  of  freedom (Fig. 1b).   The other  is  the  interaction
among  the  units, which brings about  the  differentiation.   We
believe that this emergence of recursivity or memory is a general
feature  of  coupled dynamical systems with  varying  degrees  of
freedom (e.g., the number of cells), and thus
is essential to the information and memory in biological systems.

An important consequence of our theory is global stability.
The obtained distribution of types of units (cells) is
robust against external perturbations.  Noise at the division process
may change the destiny of some individual cells, but the number distribution
of cell types is only weakly influenced by it.  Indeed this macroscopic
robustness is derived naturally from our interaction-based picture.
In the study of coupled nonlinear dynamical systems,
the stability of collective behavior is theoretically
confirmed (Kaneko 1990,92). 

Putting  the above processes into biological terms, each  unit  (
cell)  takes resources (nutrition) for reproduction  (cell
division).  First, the existence of oscillatory (biochemical) dynamics in each unit (cell)
is a natural assumption as will be discussed. The  differentiation of phases (at the first stage)  is  the
establishment of the time sharing system for resources, since the
ability to get resources generally depends on the internal  state
of  a  cell.   For example, it may be interesting to note
that different regions of the DNA replicate at different characteristic times
during the cell cycle (Alberts et al. 1983).  It is also known that the
cell cycle (of divisions) loses synchronization spontaneously 
as in our first stage, although it is often attributed to statistical
events rather than deterministic instability (Alberts et al. 1983). 
The third stage is no other than  the  division  of
labors  in  several biochemical reactions in the  cell,  since  the
differentiated units utilize different resources (chemicals). The
fourth stage where the differentiated  feature  is   epigenetically
inherited through reproduction, provides an essential  way of  maintaining
the  diversity  among  the units (cells).  Lastly, the  fifth  stage  is
simply what gives the complexity of organisms as we portrayed.
Through this process, a cooperative society of units 
emerges as a higher level \footnotemark.

\footnotetext{In this respect, isologous diversification provides
a logic for ``major transitions" (Maynard-Smith and Szathmary, 1995).}

\subsection{Cell differentiation model}

Although the isologous diversification is proposed as
a rather general scenario for biological systems, it is most important
to verify the scenario for the cell differentiation problem
through simulations of a specific model.  With this demonstration,
we provide a coherent answer to the problem raised in
\S 1.1.

First we note that it is rather natural to assume some oscillatory behaviors
in cellular chemical reactions.  Indeed, oscillations are
observed in some metabolic cycles
(Hess 1971)  as in the concentration of $Ca^{2+}$,
cyclic AMP, NADH and so on,
 while there are cyclic behaviors in cell divisions themselves,
as in the oscillations of cyclin and  MPF (M-phase-promoting factor)
(Alberts et al. 1983). 
Thus dynamical systems approach to the cell differentiation
is a rather natural postulate from physiology.
The importance of oscillatory dynamics in cellular systems has 
been pointed out by Goodwin (1963,1982).

Here we adopt autocatalytic 
reaction networks in each
cell, while interactions among cells are considered through the
medium contacting with cells. 
It should be noted that the chemicals here 
include those associated with the genetic expressions,
and even the components of DNA.   Thus our model is compatible with
the picture by a genetic switch network.  Indeed, in our simulations,
some chemicals turn to be
activated after some divisions in consistency with expressions of genes
by switchings. 
Since gene expressions are tightly linked with intracellular 
chemical reactions, which are subject to intercellular changes,
the cell differentiation satisfies the postulates of
the isologous diversification.

Previously we proposed a simple model of cell differentiation,
by including the cell division process, besides the cellular interactions.
Through simulations of this simple model, we have found
the clustering of chemical oscillations by cells at the initial
stage, and then the differentiation to rich and poor cells at
the later stage (Kaneko and Yomo 1994).  
In the present paper, we extend the  model to study
how cells are differentiated and determined successively into different types.
( see also ( Kaneko and Yomo 1995) for a brief report).

The proposed isologous diversification
theory  and  our simulation results capture the  essence  of
differentiation in the view of cellular biology.
Our result covers from the loss of totipotency,
origin of stem cells, hierarchical
organization, differences in growth rates, to the importance  of
the tiny amount of chemicals that trigger differentiation.
Some predictions are also made on the formation of tumors and their
trans-differentiation.

In our model, the units are made to interact in  a  homogeneous
environment  (well  stirred medium in biological  sense),  and hence
there  is no spatial variation.  The differentiation is  proposed
to be brought about by a dynamic mechanism, in contrast  with the (spatial)
Turing  instability.  Indeed, our dynamic scenario is  consistent
with the experimental reports on the differentiation of cells  in
a well stirred medium (E.Ko et al, 1994).  It is to be noted  that  the
authors do not disregard the spatial effect that is important  at
a later stage of development for the spatiotemporal  organization.
Indeed, some preliminary studies including a spatial factor in differentiation
suggest the validity of the present scenario 
and also the amplification of differentiation 
by spatial inhomogeneity at a later stage.

\subsection{Organization of the paper}
 
The present paper is organized as follows.
In \S 2, we present our model within a rather general framework.
Before showing the cell differentiation process, we give
a few remarks on the chemical dynamics within each cell in \S 3,
in relation with the structure of chemical network.
Explicit examples of dynamic differentiation are given in \S 4.
Following these results, we propose a general scenario of
cellular differentiation in \S 5, while some additional analysis
of the scenario is given in \S 6, based on dynamical systems theory.
We discuss the initiation of differentiation, chemical ``division of labors",
formation of tumor-like cells, and simultaneous multipled deaths.
In \S 7, some results on the numerical experiments of
cell  transplantations are given, from which the significance
of cellular memory is clarified.
\S 8 is devoted to summary and discussions.

\section{Model}

The biochemical mechanisms of cell growth and division are very much 
complicated, including a variety of catalytic reactions.  The reaction
occurs both at the inter- and intra- cellular levels.
Here we study a class of models
which captures such biochemical reaction and inter-cellular interactions.

Our model for cell society consists of

\begin{itemize}

\item { (1) Biochemical Reaction Network within each Cell : Intra-cellular 
Dynamics}

\item { (2) Interaction with Other Cells through Media: Inter-cellular  Dynamics
}

\item { (3) Cell Division}

\item { (4) Cell Death}

\end{itemize}

Our scenario to be proposed is independent of the details of modeling
as long as the items (1)-(3) are included.  For simulations,
however, we need a specific model.  Here one
example of such models is given, to propose a dynamic scenario of
cell differentiations.
The basic structure of our model is same as the previous one (Kaneko and Yomo 1994),
although the present model includes a biochemical network rather than a
simple set of reactions, to cope with the complexity in cell systems.
See Fig.3a) for schematic illustration of our modeling.

{\bf (A) Internal Reaction}

First we adopt a set of $k$ chemicals' concentrations as
dynamical variables in each cell, and also those in the medium surrounding
the cells.
Here chemicals are not specified.
They may include chemicals associated with genetic expressions,
as well as the ``metabolic" process in a very broad sense.

Based on the argument in (Kaneko and Yomo 1994), we
use the following variables; a set of concentrations of chemical substrates
$x^{(m)} _i(t)$, the concentration
of the $m$-th chemical species at the $i$-th cell, at time $t$.  The
corresponding concentration of the species in the medium is denoted as
$X^{(m)} (t)$.  We assume that the medium is
well stirred, and neglect the spatial variation of the concentration.
Furthermore we regard the chemical species
$x^{(0)}$ ( or $X^{(0)}$ in the media) as
playing the role of the source for other substrates.

The reactions $m \rightarrow \ell$ are usually catalyzed by
enzymes, which are inductive and are
again synthesized with the aids of other chemicals
$x^{(j)}$.
If this synthetic reaction is linear in $x^{(j)}$,
the concentration of the corresponding enzyme $E^{m \rightarrow \ell}_j$
obeys the dynamics
$dE^{m \rightarrow \ell}_j/dt=const.\times x^{(j)} - \delta E^{m \rightarrow \ell}_j$.
Assuming, for simplicity, fast dynamics for enzymes,
we adiabatically solve the above reaction equation of enzyme concentrations,
to get $E^{m \rightarrow \ell}_j \propto x^{(j)}$.

Let us apply the Michaels-Mentens form for
the reaction from $x^{(m)}\rightarrow x^{(\ell)}$ aided by the
enzyme $E^{m \rightarrow \ell}_j$.  Thus
the reaction from the chemical $m$ to $\ell$ aided by the chemical $j$
leads to the term $e_1 x^{(j)}_i (t)x^{(m)}_i (t)/(1+x^{(m)}_i(t)/x_M)$,
where $x_M$ is a parameter for the Michaels-Mentens' form,
and $e_1$ is the coefficient for the reaction.

Summing up, $x^{(\ell)}$ is produced with the path from the
chemical $m$, with the aid of chemical $j$.  Here $j$ and $m$ depend on
$\ell$, and generally there can be several paths for the production of $m$.
Here we use the notation $Con(m,\ell,j)$ which takes the value 1
when there is a path from the chemical $m$ to $\ell$ catalyzed 
by the chemical $j$,
and takes 0 otherwise.
In the present paper the coefficients $e_1$ and $x_M$ are identical for
all paths.

In addition, we assume that there are paths from the source chemical
and to a ``division factor".  The source is a nutrition-type chemical
for others, while the division factor includes  chemicals synthesized and
to be utilized by the division, such as the lipids for membranes, ATP, or DNA.
Here we do not
allocate it with a specific chemical, but it is assumed that there is
a threshold for the synthesis of the division factor (e.g., DNA)
to the cell division \footnotemark  as will be given in the process (C).

\footnotetext{With regards to the interplay between metabolic
reaction and the cell division factor, the present model
may have a common feature with the ``chemton model" by Ganti(1975).}

The paths from the source  chemical $x^{(0)}$ lead to the term
$S(\ell)e_0 x^{(0)}_i (t)x^{(\ell)}_i (t)$ where $S(\ell)=1$
when there is a path from 0 to $\ell$, and 0 otherwise.
The path to the final product
from some chemicals $x^{(\ell)}$,  leading to a linear decay of $x^{(\ell)}$,
with a coefficient $\gamma$.
This term is expressed by $\gamma P(\ell) x^{(\ell)}_i(t)$,
where $P(\ell)=1$ if there is a path from the chemical $\ell$ to the final
product, and otherwise $P(\ell)=0$.
Summing up
all these processes, we obtain the following contribution of the
chemical network to the growth of $x^{(\ell)}_i$ ( i.e.,
$dx^{(\ell)}_i (t)/dt$);

\begin{math}
Met^{(\ell)}_i(t)=
S(\ell) e_0 x^{(0)}_i (t)x^{(\ell)}_i (t)+
\sum_{m,j} Con(m,\ell,j)e_1 x^{(j)}_i (t)x^{(m)}_i (t)/(1+x^{(m)}_i(t)/x_M)
\end{math}

\begin{equation}
- \sum_{m',j'}Con(\ell,m',j') e_1 
x^{(\ell)}_i (t)x^{(j')}_i (t)/(1+x^{(\ell)}_i(t)/x_M)-\gamma P(\ell) x^{(\ell)}_i(t),
\end{equation}

where we note that the terms with $\sum Con(\cdots)$ represent 
paths coming into $\ell$ and out of $\ell$ respectively.
Here the chemical network can include metabolic reactions and/or
those related with genetic expressions.

The biochemical reaction here is schematically shown in Fig.3 b).
When $m=\ell$, the reaction is regarded as autocatalytic,
in the sense that there is a  positive feedback to generate the chemical $k$.
(In general, it is natural to assume that a set of chemicals
works as an autocatalytic set.)  Later we will study the case
with autocatalytic reactions only, in a more detail.

{\bf (B) Active Transport and Diffusion through Membrane}

A cell takes chemicals from the surrounding medium.
Interactions between cells, thus, occur through the medium. It is natural to
assume that the rates of chemicals transported into a cell are
proportional to their concentrations outside.
Further we assume that this transport rate also depends on
the internal state of a cell.  Since the transport here requires
energy (see e.g., Alberts et al (1983)), the transport rate depends on the
activities of a cell.  To be specific, we choose the following
form;

\begin{equation}
Transp^{(m)}_i (t) = p (\sum_{\ell=1} x^{(\ell)}_i (t)) X^{(m)} (t)
\end{equation}

The summation $(\sum_{\ell=1} x^{(\ell)}_i (t))$ is introduced here
to mean that a cell with more chemicals is more active.
Here we choose this bi-linear form for simplicity, although
nonlinear dependence on $\sum_{k=1} x^{(k)}_i (t)$ 
(i.e., with a positive feedback) leads to qualitatively similar results.
Besides the above active
transport, the chemicals spread out through the membrane with normal
diffusion by

\begin{equation}
Diff^{(m)}_i (t) = D(X^{(m)} (t) - x^{(m)}_i )
\end{equation}

Combining the processes (B) and (C),
the dynamics for $x^{(m)}_i (t)$ is given by

\begin{equation}
dx^{(0)}_i (t)/dt  = -e_0 x^{(0)}_i (t)\sum_{\ell}x^{\ell}_i(t)+
Transp^{(0)}_i (t)+Diff^{(0)}_i (t) ,
\end{equation}

\begin{equation}
dx^{(\ell)}_i (t)/dt  = Met^{(\ell)}_i(t)+ 
Transp^{(\ell)}_i (t)+Diff^{(\ell)}_i (t) ,
\end{equation}

Since the present processes are just the transportation of
chemicals through the membrane of a cell, the
sum of the chemicals must be conserved.  If the volume of
the medium is $V$ in the unit of a cell, the chemical in the medium is
diluted by this factor, and we get the following equation for the
concentration of the medium;

\begin{equation}
dX^{(m)}(t)/dt  =-(1/V)\sum_{i=1}^N \{ Transp^{(m)}_i (t) +  Diff^{(m)}_i (t) \}
-D_{out}X^{(m)}(t),
\end{equation}

where the last term corresponds to the outflow ( washout) of chemicals to
the outside of the medium. 

Since the chemicals in the medium
can be consumed with the flow to the cells, we need some flow of chemicals
(nutrition) into the medium from the outside.  Here
only the source chemical $X^0$ is supplied by a flow
into the medium. By denoting the external concentration of the chemicals
by $\overline{X^0}$ and its flow rate per volume of
the medium by $f$, the dynamics of source chemicals in the
media is written as

\begin{equation}
dX^{(0)}(t)/dt  =f(\overline{X^0} -X^0)-(1/V)\sum_{i=1}^N \{ Transp^{(0)}_i (t)
 +  Diff^{(0)}_i (t) \}.
\end{equation}

{\bf (C) Cell Division}

Through chemical processes, cells can replicate, which requires
consumption of ATP, formation of membrane, and
replication of DNA and so on.
In our model the division factor, generated 
from some chemical species, is assumed to act  
as the chemical for the cell division. Thus it is
rather natural to introduce the following condition for cell division:
The cell $i$ divides when

\begin{equation}
\int _{t_0 (i)}^T dt \sum_{\ell} \gamma P(\ell) x^{(\ell)} _i (t) > R
\end{equation}
is satisfied, where $R$ is the threshold for cell replication, and
$t_0 (i)$ is the time of the birth of the cell ( i.e., the previous division).
Here again, choices of other similar division
conditions can give qualitatively same results as those to be discussed.
The essential part for the division condition is that it satisfies
an integral form representing the consumption.

When a cell divides, two almost identical cells are formed.
The chemicals $x^{(m)}_i $ are almost equally distributed.
``Almost" here means that each cell after a division has
$(\frac{1}{2}+\epsilon)x^{(m)}_i $ and $(\frac{1}{2}-\epsilon)x^{(m)}_i $
respectively with a small ``noise" $\epsilon$, a random number with
a small amplitude, say over
 $[-10^{-3}, 10^{-3}]$.  Although the existence of imbalance is essential to
the differentiation in our model and in nature, the mechanism or
the degree of imbalance is not important for the differentiation itself.
Indeed any tiny difference
is amplified to yield a macroscopic differentiation, resulting in
the same population distribution of differentiated cells later \footnotemark.
The essence of our cell differentiation lies in the
amplification process by open chaos.

\footnotetext{Of course, the (almost) equal partition is not 
necessary for the differentiation of our simulation.
We use this partition, to stress the
intrinsic mechanism of differentiation.
By adopting unequal division, our differentiation process is accelerated 
initially.  Still, essentially the same differentiation process occurs
with this unequal partition.}

Since $x^{(m)}_i $ stands for the concentration, rather than the amount,
it might look strange to make the concentration half by division.
Here we assume that
the volume of a cell is approximated to be constant except for
a short span for the division.  During the short span for the division,
the volume gets twice and thus the concentration is made half in the above
process.  In other words, we approximately separate the stages of
the volume expansion and chemical process.
Another possible interpretation is that the biochemical reaction
process occurs within a limited region of a cell, which is not affected by
the growth of a cell size itself.

It is also possible to model the reaction process, including the
growth of the cell volume explicitly.  In this case, an additional term
for dilution is included in eqs.(4) and (5), given by 
$-x^{(m)}_i(t) \frac{dV_{cell}/dt}{V_{cell}}$, with $V_{cell}$ as the cell 
volume, which increases in proportion
to the consumption of the division factor.   When this term is included,
the division to half should be replaced by the preservation of
$x^{(m)}_i(t)$ by the division.  Indeed
we have confirmed, in several simulations,
that this modification of the model does not make any 
essential differences with regards to the qualitative behaviors.

{\bf (D) Cell Death}

In some simulations, we impose a deterministic condition for cell death.
Here we adopt the following condition for the death:

\begin{equation}
\sum_{j=1}^{k} x^{(j)}_i (t) < S,
\end{equation}

where $S$ is a threshold for ``starvation".
The choice of the death process is again rather arbitrary.  We have assumed that
a cell dies when the chemicals included therein are too little, although
a choice of similar forms is expected to give the same results.
Here, the chemicals inside the dead cells are released into
the medium.  Thus the concentration $X^{(j)} (t)$ is added
by $x^{(j)}_i/V$ at every cell death.

\section{Internal Chemical Dynamics}

Before presenting the dynamics of our cell society,  let us
briefly describe the nature of chemical (metabolic) reaction given by
eq.(1).  Roughly speaking, the dynamics strongly
depends on the number of autocatalytic paths.  Here we choose a random network
so that each chemical has a given number of outgoing
autocatalytic chemical paths.
If the number is large, only few chemicals are activated, and
all the other chemicals' concentrations vanish.  Here no other
chemical paths are active, since the ongoing reaction is just
Source $\rightarrow x^{(j)} \rightarrow$ Division Factor, without any reactions
$x^{(j)} \rightarrow x^{(\ell)}$ ( see Appendix 1).

When the number of autocatalytic paths per chemicals 
is small, on the other hand,
many chemicals are generated, but
the dynamics falls onto a fixed point without any oscillatory behavior.
In the medium number of autocatalytic paths, 
non-trivial (metabolic) reactions appear.  
Some (not necessarily all) chemicals 
are activated.  The concentrations of chemicals oscillate in time, which
often shows a switching-like behavior.  That is,
chemicals switch between low and  high values successively.
Similar behavior is also seen in randomly connected
Lotka-Volterra equations as 
saddle-connection-type dynamics (Sasa and Chawanya, 1995).

In Fig.4, we have plotted the time series of $x^{(\ell)}$
by taking only one cell and medium, without imposing the division condition
(i.e., the dynamics is given by two sets of chemicals for one cell and
the medium), where periodic alternations of dominating
chemical species are observed.  

In the present paper we discuss cases with a medium
number of autocatalytic paths, since they lead to
non-trivial (metabolic) oscillations.  
Here the term ``autocatalytic path" is not necessarily taken strictly.
Chemicals autocatalytic ``as a set" can be adopted
in the chemical network. 
See Appendix 2 for an evolutionary
account for the choice of autocatalytic networks.

\section{Example of Differentiation Process}

We have carried out several simulations of our model with 
the chemical number $k=8$, $k=16$, 
$k=32$, and $k=64$,
taking a variety of randomly chosen chemical networks with connections from
2 to 6 per chemicals.  Since typical behaviors are rather common,
we present an example
of simulation results by taking the network with $k=8$ given in Fig. 5a)
( with three randomly chosen autocatalytic paths per chemicals ).

Up to some cell numbers, all cells have identical chemical
concentrations at each instance, and oscillate synchronously.
All the cell divisions occur simultaneously, and the cell number
increases as $1,2,4,8,\cdots$.  
When the number exceeds some threshold value, the oscillation is 
desynchronized as in Fig.6, 
where the timeseries of chemicals is plotted.
In the figure, phases of oscillations of 8 cells split roughly into
2 groups.
On the other hand, 
the snapshot  values of chemicals at this stage are plotted in Fig.7
with respect to the cell index 
defined in the order of birth.
( In the example in the figure, the
differentiation starts when the cell number is 8).
At this stage, the difference by cells,
however, is seen only for snapshot values.
The average chemical concentrations over several periods
are almost identical.  

When cells further divide,
differences in chemicals start to be fixed by cells.
Average chemical concentrations measured  over periods of oscillations, 
as well as their compositions differ by cells.
The chemicals averaged from the latest division time 
are plotted in  Fig.8a)-d),
for different temporal regimes.   In Fig. 8a) ( at $t=280$),
the 3rd, 6th, 12th, and 13th cells have different chemical compositions
from others in the average.
Thus two groups start to be formed around $t=280$, while the
fixation to distinct two groups is seen around $t\approx 400$.
In  Fig. 8b, three groups are formed that is the group of
the 3,6,12,13,17,18,19,and 20th cells, and the group of 4,8,15,16,
29-32nd  cells, and that of the other cells.
For $t >600$, the distinction is much clearer as is seen in Fig 8 c)-d).  
One group of cells has a larger ability of taking the source chemical 
$x^{(0)}$, since they have larger $\sum_{j}\overline{x^{(j)}}$, with
$\overline{\cdots }$ as the temporal average.  
We call the term $\sum_{j}\overline{x^{(j)}}$ as activity. A cell with larger 
activity is called strong(er) or active here,
which divides faster than other cells.

In Fig. 9a)b) chemical oscillations of two stronger cells 
are plotted while
that of the other group is given in Fig.9c).
One can clearly see that the time series of Fig.9a) and 9c)
are different in nature, while only the phase of oscillations
differs between  Fig.9a) and 9b).  Time series of chemicals 1, 2, and 3 
overlaid for all cells are given in Fig.10 a)-c), respectively, where 
differentiation to two ( or more) groups is again discernible.
The orbits of two groups lie
in a distinct region of the phase space (see Fig.11), while
the phases of oscillations remain different by cells within each group.

Here one notes that the difference by chemicals is prominent 
for chemicals 2 and 3, while that for chemical 1 is 
much smaller. ( Chemicals 4 and 5 show similar behavior as 1).
The clearest difference is seen in chemical 3, whose concentration is
the smallest among chemicals.  As will be discussed, this suggests
the relevance of extremely dilute chemicals to differentiations.

It should be noted that the offspring of one group of cells
preserves its feature here.  In Fig. 8c), the 29,30,31,and 32nd cells
are direct offsprings of the 4,8, 15, and 16th cells.
At these later stages, differentiated features are
transmitted to daughter cells.  ( Both of divided cells
are called daughter cells throughout the paper, since there is no
principal reason to distinguish the two right after the division). 

As the cell number increases, further differentiation proceeds.
As shown in Fig.8d), each group of cells further differentiates
into two subgroups.  There are more types of cells at this stage.

\section{Proposed Scenario on Cell Differentiation}

Several simulations with a variety of chemical networks show 
similar behaviors with those given in the previous section.
In so far as we have checked, the following differentiation process
starts at some
cell number when a chosen chemical network allows for
oscillations.
Summing up these simulation results we arrive at the isologous
diversification scenario for the cell differentiation
( see Fig.2 for schematic representation).  
In our model the scenario is summarized as follows, where
each item corresponds to each stage of the isologous diversification in \S 1.2.

{\bf (1)Synchronous oscillation of chemicals and synchronized division}

Up to some number of cells, the chemical (metabolic) oscillations of all 
cells are coherent, and
they have almost same concentrations of chemicals. Accordingly,
the cells divide almost simultaneously, and the number of cells 
is the power of two.  It is interesting to note that mammalian cells 
are not differentiated up to three times
divisions.  Our result suggests that the cell differentiation is triggered 
not by a gene which counts the cell division but through
the cellular interaction.

{\bf (2) Clustering by phases of oscillations}

As cells divide further, the chemical oscillations
start to lose their synchrony.  Cells separate into several groups
with almost same phases of oscillations.  
As has been discussed (Kaneko and Yomo 1994, Kaneko 1994), this temporal 
clustering corresponds to
time sharing for resources: Since the ability to get them
depends on the chemical activities of cells, cells can get resources
successively in order with the use of difference of phase of oscillations.
Thus the competition is avoided although any control mechanism is not
imposed externally.
  
At this stage, the differentiation is not yet fixed. In other words,
only the phases of oscillations are different by cells, but  
the temporal averages of chemicals, measured over some periods 
of oscillations, are almost identical by cells.   
Cells are identical on the average.
The difference of phase, however, is a
trigger to the fixed differentiation at the next stage.

The cell number starting to show the differentiation depends on
parameters of our model.  When the nonlinearity in our model
( e.g., $e_1$ or $p$ in our model) is weak the differentiation starts
only after the cell number is large (e.g., 128), while
with stronger nonlinearity, this stage starts around the cell number 8.
For higher nonlinearity, this stage is not
clearly discerned, and the next stage starts after few divisions.

This clustering corresponds to the second stage of 
the isologous diversification, and is explained from the
studies of globally coupled dynamical systems (Kaneko,1989,90).

{\bf (3) Fixed differentiation}

After some divisions of cells ( for example, at the stage of 32 cells)
 differences in chemicals start to be fixed by cells.
The average chemical concentrations and their ratios differ by cells.
Thus cells with different chemical compositions are generated.
This leads to differentiation of cells not only with regards to
activities (i.e., differentiation between strong and weak cells) 
(Kaneko and Yomo 1994),
but also with regards to the composition of chemicals.
As is seen in Fig.8a)-d), two distinct groups
of cells are created when the cell number is 32,
and the average chemical compositions are different between the two.
( see the item (8) for the further differentiation at the later stage).

Thus we have reached the third stage of the differentiation in the
isologous diversification theory.  In connection with
the theory, the following three points should be noted:

{\bf (i) Interaction-induced change of internal dynamics}

If cells were independent, one could think that the fixed differentiation
would correspond to different attractors in the intracellular dynamics.
This is not the case.  
As will be discussed later, an ensemble consisting of only
one type of cells is unstable.  
There is an admissible range for the ratio between the numbers of
two groups of cells, which depends on the parameters of our model, and
is determined by the cellular interactions.
Thus the differentiation process depends both on intracellular dynamics
and interactions.

{\bf (ii) Two-level differentiations with phase and amplitude}

As is mentioned the phase of oscillations differs by cells even within 
each group.
Hence there are two levels of differences by cells, one for the
change of phases of oscillations, and the other for
the fixed differentiation.  Indeed this two-level differentiation gives
a source for the hierarchical organization.

It is interesting to
note that the phase difference is given by ``analogue" means, while the fixed 
difference of averages leads to a rather ``digitally" distinct separation.
Cells'  differences by phases are not rigid, since the phase diffusion
can change them:  Perturbations brought about by division are enough to shift 
the phase and destroy the memory of the previous clustering.  On the other 
hand, cells are clearly separated into few groups, distincted
by the average amplitudes  of chemicals.
This difference by the amplitude of oscillations is more rigid, since
it is not shifted continuously as in the case of the phase.  Thus 
digitally distinct groups are formed, which are stable against perturbation
such as the division.  This emergence of digital information
is the basis of the cellular memory.
A daughter of a cell of a given type
keeps its mother's character.  Indeed, the cells with stronger activities
in Fig.8, are successive daughters of a ``strong" ancestor cell, as mentioned.
(Cells 29-32 are daughters of 4,8,15, and 16, while cells 17-20 are daughters 
of 3,6,12,and 13).

The separation by the amplitude
is seen in the locus of the orbits in the phase space
of chemical values.  The orbits of the two groups of cells
lie in distinct regimes in the phase space.
As is seen in the overlaid orbits of Fig.11,
the oscillation phases are different by cells albeit
lying on the same locus in each group, while the
difference of orbits between the two groups is clearly discernible.

{\bf (iii)Separation of inherent time scales}

Another important feature here is the differentiation of the frequency.
One group of cells  oscillates faster than the other group.
Typically cells with low activities oscillate more slowly in time
with smaller amplitudes, and divide slowly, as
is seen in Fig.9 and Fig.10.  ( In Fig.10a) b), lines
taking higher concentrations correspond to stronger cells,
while the concentration of chemical 3 is smaller for stronger cells in 
Fig 10c).  Thus faster oscillations correspond to stronger cells for all
chemicals in the figures.)
Hence inherent time scales differ by cells, which
is also seen in the differentiation of speed of division.
Indeed, one group of cells divides faster than the other group of cells.
It should be noted that the inherent time scales of cells are created
spontaneously through cell divisions and differentiations \footnotemark.   

\footnotetext{
Differentiation of time scale is also studied by Volkov et al. (1992) in a
coupled oscillator model with growing degrees of elements,
corresponding to the clonal growth.}

{\bf (4) Transmission of differentiation to daughter cells}

After the above fixed differentiation, chemical compositions
of each group are inherited by their daughter cells.  
In other words, when the system enters into this stage, a cell loses
totipotency, as will be more clearly shown by the transplantation experiment 
in \S 7.  By using the term in cell biology (see e.g.,
Alberts et al. (1983)), we call
that the determination of a cell has occurred
at this stage, since daughters of one type of cells preserve the type. 
Hence a cell at this stage is called a determined one.

With the above transmission, the ``recursivity" is achieved.
In Fig.12, chemical averages of
cells between successive divisions are plotted.
Initially, chemical compositions change through divisions, but later
they come to almost fixed values. In Fig.12a) averages of chemicals
are plotted in the order of divisions.  Up to around the 90th division,
chemical averages differ by divisions, while the averages 
split into roughly two distinct groups later 
and keep the averages by divisions.
We have also plotted the ``return map" in Fig.12b), that is the 
relation between the chemical averages between the mother and daughter cells.
In the return map, the recursivity is seen as points lying
around the diagonal ($y=x$) line.  The emergence of recursivity 
is seen after some divisions.  

Note in Fig.12b) that the concentration of chemical 2 is close to zero
up to the division to 32 cells, but later is increased to keep
the recursivity.  From the molecular biology viewpoint,
this may be regarded that some genes start to be expressed to produce
some proteins. In our result here such expressions start to appear
after some divisions without any pre-programming.  

This recursivity is not expected from the studies of coupled oscillators.
For the division, we have just imposed one condition of an integral type,
which itself does not imply any recursive condition.  Through
the interference between cellular interactions and intracellular dynamics,
a cell selects an initial condition after each division, so that it keeps
its recursive structure.

It is important to note that the chemical characters are ``inherited"
just through the initial conditions of chemical concentrations after the
division, although we have not imposed 
any external mechanisms for a genetic transmission.  
(It should be mentioned, however, that we do not deny roles of
genes in the differentiation process, since our chemical can include DNA.
Our viewpoint here is neither
that genes determine everything nor that they are unimportant, but that
they are included as one of
the components in networks.)

In our model the inheritance is achieved through the transfer of ``initial 
conditions" at the division.  It should be noted that the initial chemical
composition after division is not necessarily recursive.
Indeed the return map of the snapshot chemical values after divisions
does not fall onto the fixed point as in Fig. 12b), but scatters
within some range ( although it is smaller than that at initial divisions).
This is because the phase of oscillation itself is not necessarily
relevant to keep the type.
The recursivity holds only for the averages but not for the
initial condition after each division, although it 
is attained through the choice of the latter.

Here we have reached the fourth stage of differentiation 
of the isologous diversification theory, by demonstrating
the emergence of epigenetic inheritance through
coupled nonlinear dynamics and selection of
initial conditions.
The almost ``digital" distinction of chemical characters, noted 
previously, is relevant to 
preservation of them to daughter cells, since
analogue differences of phases may  easily be disturbed by the division 
process, and cannot be transmitted to daughters robustly.

{\bf (5) Successive differentiation}

Due to the determination,
it is possible to draw a cell lineage diagram for each differentiation process,
by defining cell types by distinct chemical characters.
Generation of cell lineage gives rather useful
information to be compared with that obtained in the cell biology.
From the cell lineage, one can see how the differentiation
processes hierarchically, and how cellular memory is sustained.

In Fig.13, we have plotted the cell lineage diagram, where 
the division process with time is represented by the connected line between
mother and daughter cells.   The color in the figure shows the cell type
determined according to the chemical averages in Fig.8.
( The ``green" cell has $\overline{x^{(2)}_i}>.125$, while
the blue cell has $\overline{x^{(2)}_i}<.03$ in Fig. 8d),
with $\overline{...}$ as the temporal average.

Cellular memory is clearly seen in
this figure, where green and blue colors are preserved
through divisions for $t>400$.
We also note that the same type of determined cells (``green" cells) 
appears from different 
branches around $t\approx 500$.  
Such  convergence  of cell types from different  branches  is
also  known in cell biology.  In fact lineage analysis shows 
that in {\it C. elegans}, 
as well as in other animals, each class of differentiated  
cells,  such  as hypodermis, neuron, muscle,  and  gonad,  is
derived  from  several  founder  cells  originating  in  separate
branches of the lineage tree (Kenyon,1985).

Successive differentiation and determination of cells are
seen in the cell lineage Fig. 13. 
After two types of cells (i.e., red and green in the figure)
are differentiated around $t \approx 550$, the ``red" cells
are again differentiated into
red and blue cells. 
(see the two levels of "stronger" chemicals in Fig.8d).  Once 
this differentiation occurs, this character is fixed again, and after some
time, such characters are determined by the daughter cells.
With the cell divisions,
this hierarchical determination of cells successively continues.
For example daughters of "green" cells can later differentiate into 
``dark green" and ``light green" cells. 
Here the difference between two green-type cells
( for example that of chemicals, or the frequency)
is smaller than that between red and green cells.
(Examples of these successive determination can be
seen in Fig 8.)

Initial ``red" cells have the potentiality to
be either ``green" and ``blue" cells, while
some of the ``red" cells remain to be of the same type by the division.
Thus one can write down an automaton-like representation as
``red" $\rightarrow$ ``red", ``green", or ``blue", while
the division allows only for ``green" $\rightarrow$ ``green", and
``blue" $\rightarrow$ ``blue".  Since the ``red" cell
creates ``blue" and ``green" besides creating itself by divisions,
the red cells may be associated with stem cells \footnotemark.  When, for example
higher differentiation into
dark and light  green cells occurs, the green cell
will  again play the role of a stem cell over green-type cells.

\footnotetext{
In some simulations with a different network,
two types of cells ( with strong and weak activities) are formed,
where the division of a ``strong" cell brings about one strong cell and one
weak cell each.
In this case, the association of the strong cell with the stem cell
may be more transparent.}

With this hierarchical differentiation, successive diversification proceeds
as is postulated in the isologous diversification theory.

\section{Further Remarks on Dynamic Differentiation}

It is useful to make some remarks about how the above scenario works
and make some possible predictions on the stability of differentiation
processes.

{\bf 1)Initiation of differentiation}

In our simulation, the differentiation starts after some
divisions have occurred.  Since the division leads to almost equal
cells, a minor difference is enhanced to lead to macroscopic differentiation.
We have found that small difference of
chemicals with very low concentration leads to the amplification of
the difference in the concentration of other chemicals.
In Fig. 14, snapshot chemical concentrations are
plotted at the time step when the phase difference by cells 
is triggered.  We note that the difference of the chemical 7 ( with high
concentration) is negligible while the difference of the chemical 5 
(with very low concentration) is remarkable.  It is interesting to note that
such chemical with a low concentration is important, rather than that
with a high concentration.  This observation
reminds us of a certain protein that
is known to have a signal transmission in order to
trigger a switch of differentiation with only a small number of molecules.

The relevance of chemicals with low concentration is also
seen in the determined differentiation.  As noted in Fig.8,
the difference is most remarkable for
chemicals with low concentrations.  It is interesting to check this
proposition from experimental cell biology.  
The relevance, on the other hand,
is a consequence of our isologous diversification theory.  Since
the theory
is based on the amplification of tiny difference by a nonlinear mechanism,
difference of ``rare" chemicals by cells can be easily amplified to
lead to a macroscopic difference of cells.

{\bf 2) Specialization of a cell}

In our simulation, there are different types of a cell
as to the variety of chemicals within.
Often, only few chemicals take high concentrations
in a cell with ``stronger activity", 
while those in a ``weaker" cell are more equally
distributed.  The latter type of a cell keeps chemical variety.
In Fig.8, the difference by chemical species
is smaller in  weaker cells.
(recall the comment on Fig.10c),
where weaker cells have larger concentrations of the chemical
of tiny amount).
Generally, the bias in chemical concentrations is tended to
increase with the cell number.  This tendency is seen 
not only with regards to the number of such cells,
but also to the chemical composition of each specialized cell.

The emergence of different cell types
makes possible the division of labor in chemical reactions,
mentioned in the isologous diversification theory.
As is expected the number of cell types is increased
with the increase of chemical species $k$,
although the increase seems to be  rather slow.
The number of cell types could be much larger if
one-to-one correspondence between a chemical and a cell type 
were adopted.  Since chemicals are connected in a biochemical network
and cells originate in a single ancestor,
the number of cell types is radically reduced.

{\bf 3) Tumor type cell}

In simulations with a larger diffusion coupling, 
a peculiar type of cells appears.  They
destroy ordered use of chemical resources, which makes the cell society
disorganized.  This type of a cell is an
extreme limit of a specialized cell, and has  
a much higher concentration of one chemical than other species.
In some examples we have observed, one chemical $m$ has more 
than $10^3$ times concentration
of others.  The major biochemical (metabolic) path here is rather trivial,
since the reaction is mostly governed by
the path Source Substrate $x^{(0)}\rightarrow x^{(m)} \rightarrow$ 
Division Factor.  Chemical diversity in the cell is largely reduced.
On the other hand, the concentration  $x^{(m)}$ is so  large that
the cell divides faster than other cells.   
These cells destroy the mutual relationships among cells,
attained through the successive stages of the isologous diversification.
Taking into account this fact that these cells
destruct the chemical order sustained in the cell society,
they may be regarded as ``tumor" cells. 

The formation of these ``tumor" cells may be triggered by the mutation,
which, in our model, is represented by the ``noise" term in the division 
process ( the random number at the division process, when $\epsilon$
is larger).  Still the growth of tumor cells depends on the cellular 
interactions, for example on the diffusion constant or  the density of 
differentiated cells.  Depending on the interaction term,
errors in the division process may or may not lead to
the ``tumor"-type cells.

In Fig.15 we have plotted
the average chemicals versus the cell index, obtained
from a set of simulations with the same network as in \S 4, and
by choosing a larger diffusion coupling ($D=0.2$), and smaller threshold
for the division.  In Fig.15 b),
the ``tumor" cell starts to appear at $t \approx 140$, where the 15th cell 
(when the total cell number is 32) has a very high concentration
of the chemical 4 .  The chemical $x^{(4)}$ of the cell is around 3.6 while 
concentrations of other chemicals are close to zero.  Since the cell divides 
much faster than
other cells, its offsprings increase the number rapidly, which keep
the same chemical characters.  Thus the ``tumor" cell starts to dominate
the system. As in Fig. 15c), offsprings of the
tumor cells keep the property of an extremely high concentration of
the chemical $x^{(4)}$, although its concentration can be weaker with the 
divisions. These cells are again differentiated through divisions,
as is seen in Fig.15 d).  Here we note that chemicals other than the
species 4 are richer in the normal (weaker) cells.
Hence, the ``tumor" cell here is specialized strongly.

Of course, the extremely high concentration of {\sl one} chemical here can be
due to the simplicity of our biochemical network only with 8 species.
For a network with more components, the bias must be weakened.
However, it is still expected that there appears
a cell type with strongly biased composition of chemicals,
even if the bias is not so extreme as in the case here.
We propose that the chemical diversity is decreased in
a tumor cell in general, which is our prediction, to be
tested experimentally.

The cell lineage is given in Fig.16, where the ``tumor" cells
are plotted at the right side of the lineage, and their division
is faster.  One can also see the difference of division speeds
among the ``tumor" cells:  Some  start to lose their high activity,
and start to be differentiated.
It should be noted that the recursivity is not satisfied for
``tumor" cells.  By plotting the return map of the chemical 4 
as in Fig.12b), we have found that the
fixed point is not achieved for it, while
the return maps of other chemicals or cells of a non-tumor type
satisfy the recursivity.  This loss of recursivity implies the
heterogeneity of tumor cells. ( see also Rubin (1990)).

In our simulation, since the source chemical is limited,
the number of ``tumor" cells (with strong activity) 
cannot grow indefinitely.
With the increase of the cell density, some cells lose their
strong activity.  
In Fig. 17, we have plotted the temporal evolution of
cell numbers with three levels of activity. Around $t \approx 400$, there
is saturation of the number of ``tumor" cells, and the loss of activity occurs.

In the experiment of {\it E. coli} by Ko, Yomo, and Urabe (1994),
there appears a long-term oscillation of the numbers of cells with different
activities.  Such oscillations are observed from the longer time simulation of 
our model.  The complex dynamics of the number ratio of active to
weak cells naturally arises through  the interaction among cells.

In the example of the network Fig.5a, the ``tumor" formation is enhanced
when the diffusion $D$ is larger. In this case
cellular interactions through medium are stronger, 
and the competition for resources is more tight.
Taking also into account of the ``tumor" formation beyond some
number of cells,  we may conclude that the ``tumor" formation is
enhanced by the competition for resources by cells,
or roughly speaking by the effective density of cells.

Rubin( Yao and Rubin 1994;Chow et al., 1994, Rubin 1994a,1994b), 
in a series of experiments, has shown that
formation of a type of tumors strongly depends on the density of cells,
even if the mutation may be relevant to the trigger to it.  
The enhancement of ``tumor" 
formation in a high density, found in our results, agrees 
with the experiments.
This suggests that some type of tumor formation depends on 
epigenetical factors, and indeed
is a general consequence in an interacting cellular system.

Indeed the formation of ``tumor"-type cells is a consequence of isologous
diversification theory.  In the theory the differentiation process is not
programmed explicitly as a rule but occurs through the interaction.
Thus, when a suitable condition of the interaction is lost, for example,
by the increase of density,  ``selfish" cells destroying the 
cooperative use of resources are formed.

In the natural course of the cell differentiation, the interaction among cells
through the chemical environment is not global, but is
somehow localized in space.  Such spatial effect is important when the 
cell number and the total size of the system are increased.
If the range of interaction is limited, the effective density of cells
does not increase even if the cell number does.  Thus, the 
cell society can be developed without tumor formations.
Another way to avoid the ceaseless growth of the cellular density is
the introduction of cell death.  By increasing the cell death threshold ($S$),
the tumor formation is avoided.

{\bf 4) Cell death}

When the cell death condition is included, we have observed
simultaneous deaths of many cells.  Here we
give another set of differentiation process, by using a
different pathway given in Fig.5b), and by taking a larger threshold $S$
for cell death\footnotemark.
The chemical averages are plotted in the order of cells' indices in Fig.18,
which shows the differentiation process clearly. In Fig. 18a), snapshot 
chemical values are plotted, where slight differentiation
has started. In Fig. 18b), the fixed differentiation proceeds.  One can see
8 cells with stronger activity: 
the 31st and 32nd cells are daughters of 7th and 16th, and 33-36th cells are
daughters of 7,16,31,and 32nd, respectively. Around the time step 2000, cell 
deaths start to be observed.  Chemical averages after this stage are given in
Fig. 18c), where differentiation between strong and weak cells has occurred,
as well as slight differentiation among strong cells.
Two groups are clearly distinct as in the orbits in the phase space, say in
$(x^{(3)},x^{(6)})$.

\footnotetext{Simulation with the network of Fig.5a) and with a larger $S$
leads to the same behavior as presented here.  We use a
different pathway in order also to see the generality of our results.}

After differentiation to
three groups, cell death processes start to appear.  Through the deaths,
the number of cells varies aperiodically around 32 
as shown in Fig.19, where spontaneous deaths of multiple cells
are often observed.

The above process of cell deaths with differentiation is clearly
seen in the cell lineage diagram of Fig.20.  In Fig.20a), an early stage of
differentiation is shown.  Colors correspond to cell types, where
the activity of cells is in the order of green, blue and red for $t>800$.
One can see simultaneous deaths of cells of the same type,
arising from the same branch.  Cell deaths over longer time scales
are shown in Fig.20b) and c).
At later stage, steady distribution of cell types
is formed through the deaths.  Weaker cells exist with some ratio.

     In  the development of organisms from fertilized  eggs,
some cells deterministically die at certain stages, as termed apoptosis or
programmed death. Despite of several studies in molecular 
genetics, no genes or molecules responsible for the age of an
individual  cell have been identified, although
the presence of apoptosis-related genes had been reported in
some works. Thus, whether a
certain program or gene network governing cell aging exists or not
remains an open question.  Our simulation showing
simultaneous deaths of multiple cells as in the apoptosis,
indicates that  cell death can deterministically  occur without a special 
program.   In
other words, the fate of a cell including its death may be mainly
governed by the interaction among the cells, influencing its physiological 
state. This interaction-based apoptosis  will be
justified  by  the  transplant experiment. We predict here that
a  cell,  transplanted just before its death can survive longer  than  as
expected, indicating a change in its fate.

In the present model, no spatial structures are included.
Cells cannot move in space for richer nutritions.
Thus the number of cells is limited,
and such multiple deaths are repeated.  
With the introduction of
spatial structure, such repetition may be lost by the loss of spatial
coherence.  In our model the period
for such deaths is not fixed, but fluctuates in time.  Such fluctuation
is characteristic of an interaction-determined, rather than a genetically 
determined, system.

{\bf 5) Dependence on parameters and universailty}

We have made several simulations changing the parameters in our model.
Due to a large number of parameters, it is hard to give roles of 
parameters clearly, but they are roughly summarized as follows:

When the nonlinearity is weak, the differentiation starts later.
It seems that the differentiation will start after the
cell number is large enough, as long as there is chemical oscillation,
even if the nonlinearity is not large \footnotemark .

\footnotetext{Due to limitation of computational resources, it is 
practically difficult to check 
if the differentiation starts after the number is larger than 256
where the computation requires much longer time.}

Increasing the parameter $p$ or $\gamma$
enhances differentiation.  These changes increase the
amplitude and/or the frequency of oscillation, and  may be regarded as
the increase of ``nonlinearity".  The decrease of the growth threshold
$R$, on the other hand, suppresses the differentiation.  Cells remain 
identical, although chemical speciation within is often enhanced, and
only few chemicals' concentrations are larger than zero;
in other words an ensemble of ``tumor"-type cells 
tend to be formed.
Increase of the diffusion $D$ has a similar effect as the decrease of $R$.

As mentioned, single cell dynamics depends on
the network, which reflects on the behavior of cellular society.
Generally speaking, our differentiation scenario
works as long as a single cellular dynamics 
allows for the oscillatory dynamics.
As for paths to the division factor ($P(\ell)$) and paths from 
the source chemical ($S(\ell)$),
there is a tendency that the differentiation process is
enhanced when these connections are not full ($=k$).

{\bf 6) Macroscopic stability}

To close the section, it should be noted that our scenario,
although based on chaotic instability, is rather robust against 
changes of initial conditions or errors in the division process.  
Of course, which cell becomes
one given type can depend on the initial conditions.  
On the other hand, 
the number distribution
of each type of cells, as well as the cell lineage diagram,
does not depend on initial conditions, as long as very special
initial conditions are not adopted,
as in the transplantation experiment ( see the next section).

The variety of cell types and their number distribution are
robust against the noise (error) in the 
division process ( which may be regarded as the
mutation when the corresponding chemical is DNA).
On the other hand, when the population of one type of cells
is decreased (e.g., by external removal),
the distribution is recovered through further divisions.

This kind of robustness at an ensemble level, indeed is
expected from our isologous diversification theory,
since the stability of macroscopic characteristics is
attained in coupled dynamical systems ( Kaneko 1992,94;
Yomo and Kaneko 1994).  This robustness gives a key to understand how a stable
cell society is formed, without being damaged by somatic mutations.

\section{Transplantation Experiments and Cellular Memory}

The differentiation in our isologous diversification
scenario originates in the interaction among
cells, but later, at the third stage, chemical characters of a cell are
memorized through the initial condition after division.
The differentiation at the former mechanism is 
reversible, while the latter mechanism leads to determination.
It is interesting to note that the determination is not
implemented in the model in advance, but emerges spontaneously at some 
stage of cell numbers.

In the natural course of differentiation and in the simulations in \S 4,
however, it is not possible to separate the memory in
the inherited initial condition from the interaction with other cells.
To see the tolerance of the memory in the  inherited
conditions, one effective method is
to choose a determined cell and transplant it to
a variety of surrounding cells, that are not seen in
the ``normal" course of differentiation and development.
Let us discuss the results of these ``transplantation" experiments.

In real biological experiments on differentiation,
some ``artificial" initial conditions are adopted by
the transplantation of some types of cells.
To check the validity of our scenario and to see the
tolerance of the memory in the inherited initial condition,
we have made several 
numerical experiments taking such  ``artificial"
initial conditions. Here we have made the following observations,
by initially taking cells obtained at the normal diffusion process
and putting them into undifferentiated cells
at an earlier stage.  

\vspace{.1in}

{\sl i) starting only from few differentiated cells of the same type
in addition to undifferentiated cells  }

Fig. 21a) gives the evolution of average chemical concentrations by cells  
starting from 4 determined cells ( whose cell index is from 1 to 4) and
4 undifferentiated cells.  The former cells are sampled 
from later stages ($t=691$) in the simulation of \S 3, while the latter 
ones come from the former stages ($t=205$),
In Fig.21a), 14,15,16,17th cells are the first daughters of 1,2,3,4th cells
and the cells from 25 to 32 are the daughters of the above 8 cells.
The former group of cells
keeps the type, whose offsprings remain the same type.  Thus 
the determination is preserved, and the memory in the
inherited initial conditions is robust against the change of cell
interactions.  The undifferentiated cells,
on the other hand, start to differentiate
to form many types of cells, as is seen in Fig.21a).

This robustness of cell memory is kept as long as the ratio of initial 
determine cells to undifferentiated cells is not too much.  
( see iii) for the case otherwise).

\vspace{.1in}

{\sl ii) starting  from the mixture of different determined cells }

Again the cell memory is preserved, and daughters of a cell keep
the same character.
In Fig. 21b),
two types of determined cells are initially taken,
7 cells for one type ( with the  index from 1 to 7), and 1 cell 
( the 8 th cell in the figure) of another type, obtained from \S 3 at $t=691$.  
The Fig.21b) shows the average chemical pattern
after twice divisions.  The 16th, 31st, and 32nd cells, which are offsprings of
the 8th cell keep the character, while other cells remain to be the other type.
Some other simulations also show that cellular memory is 
preserved as long as initial distribution of cells
is not dominated by only one type of cells
( see iii) for the case otherwise).

\vspace{.1in}

{\sl iii) starting only from few differentiated cells of the same type}

In most cases, these cells start to dedifferentiate again to generate
different types of cells.  Some of them keep their character
while others ( and their offsprings) become a different type.
If initial  distribution of cells is dominated by one type of determined
cells in cases i) and ii), again some of them start to dedifferentiate
and become a different type of cells.  In Fig.21c), the average chemical 
pattern is plotted starting from 20 cells of one type of determined cell
of \S 3 at $t=691$.  Cells with the index from 1 to 16 and 19
preserve their character, while 17th and 18th cells become a different type,
and the 20th cell is transdifferentiated to another type.
Here their offsprings again keep the character:
22nd and 23rd cells are daughters of 17th and 18th, while the 21st cell is
a daughter of 20th,
and 41-42nd are those of 20th and 21st.  Thus determination again occurs
after the process of dedifferentiation.

In an example with a different chemical
network (with a larger number (4) of connections), 
we have found the formation of ``tumor"
cells with ongoing simple chemical reaction paths.
Again, these cells lose the variety of chemicals and
destroy the ordered use of resources.

Summing up i)-iii),
we can conclude that the cell memory is preserved mainly in each cell, 
but cellular interactions are also important to sustain it.
The achieved recursivity in \S 4 is understood as the choice of 
internal dynamics through cellular interactions.  

Thus the cellular interactions play the role not only of the
trigger to differentiation, but also of the maintenance of diversity
of cells.   Internal cellular memory is maintained as long as
the diversity is sustained.  The relevance of interactions to diversification
is a key concept in our isologous diversification.

\vspace{.1in}

{\sl iv) differentiation of ``tumor" cells}

Another interesting initial condition is the use of
``tumor" cells.  Starting only from tumor cells, their offsprings
remain to be the same type initially.  As the divisions are repeated,
some cells' activities get weaker and start to be differentiated.
This is a consequence of our theory that is strongly based
on cellular interactions.
 
Such differentiation of ``tumor" cells
is promoted by adding undifferentiated cells initially.
As an extreme case of ``tumor" cell, let us
consider a cell with $x^{(m)}$ is large
and $x^{(\ell)}= 0$ for $\ell \neq m$.  Such cell
can divide faster if there are paths from the source to $m$ and 
from $m$ to the division factor.
When there are a large number of autocatalytic 
paths, $x^{(\ell)}$ remains to be zero for the cell, whose offsprings keep
the same type.  In this extreme case,  the cellular society remains to 
consist only of ``tumor" cells.
Even in this case it is found that ``tumor" cells are differentiated by
adding undifferentiated cells ( taken at the initial stage
of the simulation of our model).

\section{Discussions}

\subsection{Summary and biological relevance}

To sum up we have proposed isologous diversification theory
on cell differentiation, by introducing a model
based on the interacting cells with chemical oscillations and the
clustering of coupled oscillator systems.
From the simulation of the model, 
we have observed 
successive spontaneous differentiations and their transfer 
to daughter cells, without any external mechanism.

Let us summarize the consequences of our simulations.

\begin{itemize}

\item
Cell differentiation occurs through the interplay between 
intracellular chemical reaction dynamics and
the interaction among cells through chemical media.

\item
Cell differentiation is initiated by the clustering of chemical oscillations,
appearing at some cell number, which is explained by
general features of coupled nonlinear oscillators.

\item
Chemicals with tiny amounts in cells are relevant to a trigger to differentiations.

\item
With further divisions cells lose totipotency, and offsprings preserve
the same character.  This recursive division of cells appears only
after some stage of cell divisions.

\item
Distinct and memorized cell types are formed by the clustering of
the amplitude, rather than the phase, of oscillations, which leads to
the emergence of digital changes in chemical concentrations.

\item
Determined cell changes are characterized by
the cell activity and chemical compositions.

\item
Inherent time scales, given by the oscillation period,
differentiate by cells.  Generally active cells oscillate faster,
and divide faster.
This separation of time scale brings about
the separation of growth speed of cells, and leads to the disparity between
rapidly growing and inactive cells.

\item
Generally speaking,
cells whose chemicals are concentrated on few species, are stronger in
activities and divide faster.  Cells keeping a variety of chemicals
divide slower.  There is a negative correlation between the growth speed
and the chemical variety within a cell.

\item
Successive differentiation appears at a later stage, which leads to
cells that bring about only a range of cell types successively.

\item
Determined cell types formed at the later stage are
preserved by their transplantation to other cell society as long as
there are not too much cells of the same type.

\item
It is possible to dedifferentiate cells, by putting them in
some conditions such as overcrowded cells of an identical type.

\item
Spontaneous multipled deaths appear through the interaction of cells,
after cells are differentiated.

\item
A type of tumor-like cells is formed,  depending on the cellular
interaction.  These cells destroy the ordered use of resources
attained in the cell society.  For example the formation is
enhanced by the cell density or the diffusion coupling.
Transplantation of cells of the same type may enhance the tumor formation.

\item
Such type of tumor cells can be differentiated to normal cells, 
through the interaction
with other cells.  The differentiation can be enhanced by
adding undifferentiated cells.

\end{itemize}

It is interesting to note
that the above picture is consistent with
a variety of experimental results
such as loss of totipotency,
origin of stem cells, hierarchical organization of differentiation,
separation of growth speeds by differentiation, tumor formation,
importance of tiny amount of chemicals
for the trigger of differentiation,
and so on.  It should be mentioned that these
results naturally appear as a general consequence of
our isologous diversification theory without pre-programmed implementation,
and are independent of
detailed modeling. We should also mention that our theory is compatible with
the genetic switching mechanism for the differentiation.  
Here such switching-type expression appears naturally
through cellular interactions.

   As is  described in \S 6.2, once cell differentiation  in  our 
model  reaches the fifth stage in which the  cells are successively 
differentiated,  the variety of components in each cell is  negatively  
correlated to the growth rate or activity.  That is,  a 
cell  with a higher division rate tends to have a lower variety  of 
metabolites or simply saying, these cells have a simpler metabolic network.  
This proposition can possibly be examined by cell 
biologists.   For instance, in the process of development of  the 
organism, one can sample existing various types  of 
cells in several stages of successive  differentiation.  Then it
is possible to  
determine the variety of the  components  including 
macromolecules and the growth rate for each cell type and check
the correlation between the two.  Thus, the authenticity of  the 
isologous  diversification  theory can easily be  tested  in  the 
laboratory  scale.  

Furthermore, the theory can be  extended  for 
medical application.  As mentioned in \S 6.4, the tumor cells in our 
model  are in the extreme case of the loss of the variety, where the  cells 
lose some of the metabolites or become to have a simpler network 
to  achieve  the faster growth rate.  One way to bring  a  tumor 
cell back to normal is to supply the metabolites which they  lost 
through  the development.  The cell will then recover its  normal 
network  and hence, will grow harmoniously with  the  surrounding 
cells.   Similarly,  in  order for cancer  cells to regain  the 
normal physiological state even with some mutations on their DNA,  
they  are fused with the liposomes encapsulating the  cytosol  of 
normal cells or undifferentiated cells, which are expected to 
include some of metabolites or macromolecules that the cancers lack.

\subsection{Isologous diversification}

It should be noted that the introduction of tiny difference at the division
is not essential to our differentiation scenario.  A system 
composed of identical cell states is unstable when
the interaction is strong.  
Even if parameters or initial conditions of a cell are
different, they may not be essential to the differentiation.
Indeed in the clustering of coupled dynamical systems,
it is known that an element with a different parameter
can oscillate almost coherently with others, while 
elements with identical parameters split to two ( or more) groups 
(Kaneko 1994). Here the parameter variation of elements  is not essential to
the grouping of them.
Thus it can happen that a cell with rather different
parameters or initial conditions remains to be of the same type with other 
cells, while some
cells with identical parameters or close initial conditions
become a different type.
In this perspective, it is expected most somatic mutations are
irrelevant to differentiation, unless it brings about
a drastic change in the parameter for the interaction and intracellular 
dynamics.

Our proposal here is that the differentiation and diversification
are not due to variations by reproducing errors but by the dynamical
instability.  This is the reason why we
 have called our theory ``isologous diversification", to stress
the inherent tendency of differentiation of identical elements.

Indeed we believe this ``isologous diversification" can be generally applied to
a variety of biological systems, because it
is based on our study of coupled dynamical systems,
which is expected to be universal in a class of interacting,
reproducing, and oscillatory units.
In particular we have succeeded in showing a mechanism of
division of labors through differentiation and
segregation into active and inactive groups.
Since the picture is based on coupled dynamical systems
theory, it is expected to be
applied to economics and sociology, which enables us
to discuss the origin of differentiation, diversity, and complexity there.

In biology, the origin of multicellular organism
is directly related with the above picture and our result here. 
For its origin, some mechanism of differentiation of 
identical cells is necessary which leads to divisions of labor, while
the differentiation reaches at the stage that only
one group of cells brings about its offsprings.
According to our results, this feature of a
multicellular organism spontaneously emerges as
a consequence of strongly coupled reproducing units.
It is not a product of chance, but of the necessity in the course
of the evolution, since reproducing units should 
reach a strong coupling regime by their growth.
It should be noted that our study explains not only the origin,
but also the maintenance of diversity ( see also Kaneko and Ikegami (1992)
for the relevance of weak chaos to the maintenance).

\subsection{Some future problems}

Our theory of differentiation raises some basic problems.
To close the paper we discuss five of them, the first two of which are
related with the extension of our modeling, while the latter three
are of more fundamental issue.

{\bf Universality and simpler models}

Our results are rather universally observed as long as
individual dynamics allows for some oscillations. 
Since globally coupled dynamical systems are known to show
spontaneous differentiation as the clustering (Kaneko 1989,90),
we may expect that our differentiation is 
universally observed in a large class of a coupled system of
nonlinear reproducing units.

To search for a simpler model, we have also checked a model only with a 
phase variable (Kaneko 1994).
So far this model shows  the stage of phase clustering and can explain
its relevance to time sharing for resources, but cannot
show the stage of the disparity and fixed differentiation.
For the stage, a model with amplitude clustering is required, as in ours.

Note that the clusterings in our model occur in a dual space,
that is, in chemical species and in the cell index.
Indeed one can construct a coupled map model with
dual space, which shows the clustering in cell index and/or chemical
species.  Here the 
clusterings between cell indices and chemical species compete with each
other.  Construction of minimal models with the differentiation process
will be an interesting problem as dynamical systems theory.

{\bf Introduction of spatially local interaction and development}

In the paper we have assumed that chemical medium
is well stirred, and all cells interact with all others
uniformly through the medium.   In the developmental process
of a multicellular organism, spatially local interactions among cells
are, of course, important, as the development proceeds.

We have made some preliminary simulations including spatial
inhomogenization of the medium.  So far the result shows that
the differentiation process starts in the same manner as
that presented here.  First, the
phase of oscillation is differentiated according to
its division.  At a later stage, cells close to each other
start to be differentiated following the
scenario in the present paper.
Then a cell's character is fixed, depending also on
the locality in space.  At later stage, due to the
local interaction, spatial organization of differentiated
cells occurs, leading to the pattern formation,
as in the pioneering study by Turing.

Our proposal along the present paper is that the temporal
organization of cells occurs first, leading to
cell differentiation, and later the pattern formation
follows.  Hence we have focused on the
global interaction case here, although, of course
the spatial organization is the next important issue,
as will be studied in future.

Since distant cells do not interact directly with each other,
differentiation as well as its determination is often enhanced.
Another consequence of the spatial separation is the
suppression of competition for chemical resources,
which makes the simultaneous cell deaths smaller in number, and
localized in space.

It will be of interest to include the cell motility following an
intercellular force, to study cellular rearrangements  
leading to the pattern formation.
This, for example, may result in  
a simple model for the differentiation process of 
Dictyostellum discoideum.

Another extension of our model is
the use of a ``batch-type" simulation where chemical resources
do not flow into the media but are kept constant.
Indeed, there is no flow of chemical resources
from the outside, during the early developmental process of
an egg such as the sea urchin.  Our model
can directly be extended to this ``batch" type simulation,
by cutting the flow from the outside, i.e., by
setting $D_{out}=f=0$ and taking a higher density of
nutrition initially.  Since the nonlinear dynamics and the interaction
are still included, it is expected that the
differentiation process follows the stages of the present theory,
as long as the initial nutrition is sufficient.
Here, the final number of cells depends on the initial
amount of nutrition, while, the distribution of the cell types
should be almost independent of it
as long as it is not too small.
This robustness may give a prototype of the
developmental stability
found in Driesch's experiment on the sea urchin.

{\bf Cellular memory}

The emergence of the cell memory, found in our system,
raises an important issue in coupled dynamical systems.
Is the memory stored in
each cell or in an ensemble of interacting cells?  Our proposal here
is that it is preserved through intra-inter-dynamics,
that is partly within each cell, and partly distributed
in the cell society.
The existence of multiple cellular types can be related with 
coexisting attractors corresponding to different basins
for initial conditions, while the stability of differentiation
is sustained by the interaction.  Indeed, with the interaction,
the distribution of cell types is almost independent of initial conditions,
and is also robust against perturbations such as removal of some cells,
or other possible environmental changes.

This is a novel form of memory in
dynamical systems.  Due to the interplay between
intracellular dynamics and interactions, the fixation of memory and
diversification are compatible.
It is important to clarify the condition of the emergence of cell memory,
as well as to search for applications of this type of memory to other biological
systems such as immune or neural networks.

{\bf Recursivity through choice of initial conditions}

The next problem is the initial condition selection with recursivity.
As several divisions proceed, each cell enters into the stage
whose daughter keeps the same character.  It is
recursive in the sense that the initial condition of a cellular
state after a given division leads to the next initial condition
after the division so that it has the same cellular character.
With this sequence of initial conditions
some condition must be satisfied to keep the same character.
For this, some chemicals should remain at some range, although not
necessarily are completely identical.
We note that the initial condition itself after each division
does not fall on to a fixed point.  The phase of oscillations
at each division is rather arbitrary.  The recursivity is
achieved as a fixed point of the average motion as given in Fig. 12.

A novel framework is required to discuss the stability at the average level,
and the selection mechanism of initial conditions so that the system is
recursive.  To be
recursive, a set of initial chemicals should be determined rather precisely
while others are loosely determined.  In our problem,
this choice is also dependent on the environment
(medium), which depends on other cells' states.
The formation of tumor cells is understood as the loss of
recursivity, in this context.
Detailed discussions on this initial condition problem will be
discussed elsewhere, where
the problem of separation of egg (DNA) and chicken (protein)
will be reconsidered along this line.

{\bf  Open chaos}

Besides this viewpoint of coupled dynamical systems, it should
be noted that our system is ``open-ended" in the sense that
the degrees of freedom increase with the cell division, where
the notion of ``open chaos" (Kaneko 1994b) will be
useful to analyze the mechanism of cell differentiation problems.

The last, but important question is the evolution of
metabolic network.  In the present paper, we
have chosen randomly connected autocatalytic networks. Even among
the random networks,
the dynamic behavior depends
on the topology of the network, as well as
the number of autocatalytic paths, which is most relevant.
The metabolic network in a cell is constructed
through the evolution, and differs from that
constructed as a random graph.  The network is
history dependent, and is constrained by the
survivability within a cell society.
Evolution of metabolic pathways within the cellular interactions
and intracellular dynamics should be studied in future.

{\sl acknowledgements}

The authors are grateful to  T.Ikegami, S. Sasa, N. Nakagawa, T. Yamamoto,
and I. Urabe for stimulating discussions.
The work is partially supported by
Grant-in-Aids for Scientific
Research from the Ministry of Education, Science, and Culture
of Japan.
The authors would like to thank Chris Langton for their hospitality
during their stay at Santa Fe Institute.

\section{ Appendix 1: Winner-takes-all mechanism in chemical reaction dynamics}

In this appendix, we briefly discuss the chemical reaction dynamics of
our model, for a single cell.

When there is a two-way connection between chemical species,
the winner-takes-all mechanism can be expected.
This can be understood by taking a simple example with two chemical
species:

\begin{math}
dx^{(1)}/dt=x^{(2)}x^{(1)}/(1+x^{(2)})-x^{(1)}x^{(2)}/(1+x^{(1)})
\end{math}

\begin{math}
dx^{(2)}/dt=x^{(1)}x^{(2)}/(1+x^{(1)})-x^{(2)}x^{(1)}/(1+x^{(2)})
\end{math}

As is easily seen from this equation, the difference
$x^{(1)}-x^{(2)}$ is amplified with time,
and goes to a state with either $x^{(1)}=0$ or $x^{(2)}=0$.
Thus, the bidirectional connection tends to lead to
competition of chemical species in the present model.  Indeed
selection of few chemicals is seen when there are
many bidirectional pathways.

As the extreme limit, let us consider a case
with full connections of paths.  In this case, we have observed that
as an attractor only one chemical species has a finite value,
and others vanish in a strong nonlinearity regime ( i.e., with
large $e_1$).  Here the dynamical process is just the
selection of one chemical species through the competition for
resources.

\section{ Appendix 2 : Choice of Internal chemical dynamics}

When the number of autocatalytic paths is large, the
mechanism mentioned in the Appendix 1 works, and 
only one or few chemicals are activated.
When the number of autocatalytic paths is small, on the other hand,
many chemicals are generated, but
the dynamics is stabilized and goes to a fixed point state.
In the medium number of autocatalytic paths, 
non-trivial metabolic reactions appear as mentioned in \S 2 \footnotemark.
Periodic alternations of dominating
chemical species are observed.  
Depending on the nature of connections, we have seen several types
of oscillations, although chaotic ones are not found so often.
When the number of chemicals is larger,
the alternations are more complicated as in Fig 22 a)b) where the dynamics is
possibly aperiodic.

\footnotetext{
Besides the number of autocatalytic connections,
there is further dependence on each pathway of the network.
We have examined several random networks of 2 autocatalytic 
paths per chemicals for $k=8$. 
Some of the networks
lead to oscillatory dynamics, while others show
fixed point dynamics with few chemicals of high concentrations,
although the number of autocatalytic paths is identical.
So far we have not found a simple criterion for the oscillatory behavior.}

If there are a few non-autocatalytic (i.e., $Con(m,j,\ell)$ with $j \neq \ell$)
paths, fixed-point states are stabilized, and oscillations are hardly observed.
We adopt the intracellular dynamics consisting only of the 
autocatalytic paths
whose number is medium per chemicals (from 2 to 4),
since they provide examples with
ongoing non-trivial metabolic reactions. 

There is indeed a reason for this choice from an evolutionary point of view.
In the evolutionary process of metabolic reactions, novel chemicals
are successively included in the network.  Let us consider the inclusion 
process
of a new chemical $J$.  Its chemical concentration must be amplified
through the chemical network process, otherwise,
it is diluted and disappears through divisions.  
Since the new chemical $J$ did not exist before, 
$dx^{J}/dt=0$ if $x^{J}=0$.  On the other hand,
for the growth of the concentration of 
chemical $J$ in its presence, $dx^{J}/dt>0$ must fold for $x^J >0$.  Hence the
condition $\frac{\partial}{\partial x^{J}}\frac{dx^{J}}{dt})>0$.
Thus it is expected that 
$\frac{dx^{J}}{dt} \propto (x^{(J)})^{\alpha} $ with $\alpha >0$.
Thus, some kind of autocatalytic processes for the chemical $J$ must exist.
In this way, it is expected that chemicals with autocatalytic 
processes are adopted successively, through the evolution of metabolic process.

As mentioned in \S 2 the term
``autocatalytic path" is not necessarily taken strictly, but
may assumed to represent chemicals autocatalytic ``as a set" 
( see Fig.23 schematically).  
In such case, one may approximately represent the set of chemicals
by one variable  $x^{(\ell)}$, and adopt an autocatalytic reaction for
$x^{(\ell)}$.  Thus our chemical reaction may be interpreted to
represent the network composed
of a set of autocatalytic networks, expected from the
evolutionary process.

\pagebreak

Figure Caption

\vspace{.1in}

Fig.1  Schematic representation for two pictures of
cell differentiation.  a): fixed landscape.  b): Our picture
based on the interplay between intra- and inter- cellular dynamics.

\vspace{.1in}
Fig.2
Schematic representation of our isologous diversification.
Each spiral represents oscillatory dynamics.
In the figure each stage shifts to the next by a single 
reproduction process (e.g., a cell division)
for simplicity, but in general there are several reproductions
in each stage.
See \S 5 for details.

\vspace{.1in}

Fig.3  Schematic representation of our model: a) the whole
dynamics of our system; b) 
chemical reaction within each cell.

\vspace{.1in}

Fig.4   Overlaid time series of $x^{(m)} (t)$ of a
single cell in medium, obtained from a network with three connections
of 8 chemicals whose connection is given in Fig. 5a).
Each line with the number $m=$1,2,4,5,7,8 gives the timeseries of
the corresponding chemical  $x^{(m)} (t)$.  Note that the chemical 2 has a lower
concentration and appears only around the bottom of the figure,
while the concentrations of chemicals 3 and 6 are too low
to be discernible in the figure.
The parameters are set as
$p=10.0$, $e_0=e_1=1$, $\overline{X_0}=40$, $\gamma=0.2$, $x_M=10.0$,
$D_{out}=f=0.005$, and $V=1000$, while the division and death processes are not
included.

\vspace{.1in}

Fig.5 
Biochemical network adopted in the simulations shown in the present paper.
(a)3 outgoing connections per chemical   (b) 2 outgoing connections per chemical.
The species with a double circle has an arrow to the division factor
product ($P(\ell)=1$), while all chemicals have arrow from the source 
chemical 0 (i.e., $S(m)=1$ for all $m$).

\vspace{.1in}

Fig.6
Time series of $x^{(4)}_i (t)$ for $134<t<140$, overlaid over all cells.
Cell division occurs around $t=137.8$, when the cell number
doubles from 8 to 16.
The cell index is defined in the order that the cell is born.
In the simulations of the present section
(given in Fig.6--14,and 22) we adopt the chemical 
network of Fig.5a) and  use the parameters
$p=10.0$, $e_0=e_1=1$, $\overline{X_0}=40$, $D=0.02$, $\gamma=0.2$, $x_M=10.0$
$R=2000$,  $S=0.05$, 
$D_{out}=f=0.005$, and $V=1000$.

\vspace{.1in}

Fig.7 

Snapshot chemical concentrations of $x^{(m)}_i(t)$ at $t=114$,
when the cell number is 8.  The concentration of chemicals
 $x^{(m)}_i$ for $m=1,4,$ and $8$ are shown, since the concentration 
of other chemicals are very low at this time instance.
Lines connecting  $x^{(m)}_i$ are just for the sake of presentation.
As described in the text, cell's index is labeled in the order of the birth:
when the first division occurs, one of the cells remains to be
denoted as 1, while the other is labeled as 2. When the next division occurs
for the cell 2, for example, one of the divided cells has the cell index 3,
while the other remains to be 2, and so forth.
Data for Fig.6 -14 are obtained from a
simulation  with parameters given in Fig.6. 

\vspace{.1in}

 Fig.8 

Average chemical concentrations of $x^{(m)}_i$.  
The average
is taken over the time steps from the latest cell division
(or since its birth, when it has not experienced a division yet).
Each color corresponds to each chemical $\overline{x^{(m)}_i}$ for 
$m=1,2,\cdots 8$, with $\overline{...}$ as the average.
Note that the line is plotted only for visualization, and the
values at integer cell indices give corresponding $x^{(m)}(i)$.
The concentration of the chemical 6 vanishes after some time,
and is not plotted.
(a) $t=280$, when the cell number $N=16$,
(b) $t=400 $ and $N=32$
(c) $t=600$ and $N=32$
(d) $t=940$ and $N=64$.

\vspace{.1in}
Fig.9:
Time series of $x^{(m)}_i(t)$ for $800<t<805$, overlaid over all chemical
species $m$. (Each line corresponds to each chemical).
(a) for the cell $i=2$ (b) for the cell $i=3$
(c) for the cell $i=4$.

\vspace{.1in}
Fig.10

Time series of $x^{(m)}_i(t)$ for $800<t<805$, overlaid over all cells $i$.
Each line corresponds to the time series of each cell.
(a) for the chemical species $m=1$ (b) for the chemical species $m=2$
(c) for the chemical species $m=3$.
In (a) and (b), oscillations with a larger amplitude correspond to cells
with larger activity ($\sum_m \overline{x^{(m)}_i}$), while
they take smaller value for the chemical 3 given in (c).

\vspace{.1in}

Fig.11 

Orbits of chemical oscillations.  Plotted are
$(x^{(5)}_i(t),x^{(8)}_i(t))$ for $ 800<t<900$ overlaid over all cells
(whose number is 64).
Two groups are clearly seen.

\vspace{.1in}

Fig.12 

(a)Chemical concentrations $\overline{x^{(m)}_i}$, averaged
over two successive divisions, are plotted in the order of divisions.
The upper column shows the expansion of the lower for divisions later than 60th.
Different marks correspond to different chemicals, while
lines are plotted only for convenience.

(b) Return map of chemical concentrations $\overline{x^{(m)}_i}$ averaged
over two successive divisions.  A daughter cell's average concentration
is plotted versus its mother's cell's average before the division to 
the daughter. Chemicals 2,5, and 8 are plotted with different marks,
while the dotted lines are drawn only for convenience.
The lower column is the expansion of the upper column.

\vspace{.1in}
Fig.13

Cell lineage diagram corresponding to the simulation in Fig.6--12.
The vertical axis shows the time,
while the horizontal axis shows a cell index.  ( Only for
practical purpose of keeping track of the branching tree, we define
the index for the lineage 
as follows: when a daughter cell $j$ is born from a cell $i$'s
$k$-th division, the value $s_j =s_i + 2^{-k}$ is attached to the cell $j$
from the mother cell's $s_i$.
The cell index for the cell $j$ is the order of $s_j$,
sorted with the increasing order.  Note that the idex for the lineage diagram
is different from the cell index adopted in other figures, where
the index is given just as the order of birth).
In the diagram, the horizontal line shows the division from the
cell with index $n_i$ to $n_j$, while the vertical line
is drawn as long as the cell exists (until it dies out).
Color corresponds to the cell's character defined from
the average chemical pattern.  After differentiation
the activity of a cell is in the order of green, red, and blue, while
initial red cells correspond to undifferentiated ones.
The ``green" cell has $\overline{x^{(2)}_i}>.125$, while
the blue cell has $\overline{x^{(2)}_i}<.03$ in Fig. 8d).

\vspace{.1in}
Fig.14 

Snapshot chemical concentrations of $x^{(m)}_i$, at $t=63$,
just the onset of chemical difference by cells (clustering).
Chemicals 4,5, and 7 are plotted in a logarithmic scale,
in the order of cell index, the order that the cell is born.

\vspace{.1in}
Fig.15

Average chemical concentrations of $x^{(m)}_i$.  
The average is taken over the time steps from the latest cell division
(or since its birth, when it has not experienced the division yet).
We adopt the chemical 
network of Fig.5a) and  use the parameters as for Fig.6, except
$D$ and $R$, which are taken to be $D=0.2$ and $R=500$.
(a) $t=130$, when the cell number $N=32$,
(b) $t=140$ and $N=32$
(c) $t=380$ and $N=119$
(d) blowup of (c), with the blowup of the vertical axis.

\vspace{.1in}
Fig.16

Cell lineage diagram corresponding to the simulation of Fig.15.
The vertical axis shows the time,
while the horizontal axis shows a cell index, defined as
in Fig.13.

\vspace{.1in}
Fig.17  

Temporal change of the number of cells $N$, and 
very strong, strong, and weak cells for the simulations of
Fig.15 and 16.  They are defined
by $\sum_m x^{(m)} _i$ $>10$, $1<\sum_m x^{(m)} _i<10$,
and $\sum_m x^{(m)}_i<1$,respectively.

\vspace{.1in}
Fig.18

Snapshot (a) and average (b,c) of chemical concentrations of $x^{(m)}_i$.  
In the simulations for Figs. 18--21, we adopt the chemical 
network of Fig.5b) and  use the parameters
$p=10.0$, $e_0=e_1=1$, $\overline{X_0}=10$, $D=0.02$, $\gamma=0.2$,
$x_M=10.0$, $R=100$,  $S=0.01$, 
$D_{out}=f=0.005$, and $V=1000$.
For $t>700$, concentrations of chemicals 4 and 8 almost vanish,
which are not plotted in the figure.
The values at (integer) cell indices give corresponding chemical concentrations
while lines are drawn  only for the clarity of the figure.
The circle or square is plotted for $x^{1}(i)$ to show the
existing cell clearly.
(a) Snapshot: $t=717$, when the cell number $N=32$,
(b) Average over $718<t<1822$ and $N=64$
(c) Average over $2355<t<2962$; Those cells are dead whose
indices do not have corresponding circles for $x^{1}(i)$
(such as the cells between 7 and 15 or between 37 and 64).

\vspace{.1in}

Fig.19  

Time series of the number of cells $N$,
for the simulation of Fig.18.

\vspace{.1in}

Fig.20

Cell lineage diagram corresponding to the simulation in Fig.18--19.
The diagram is plotted in the same manner as Fig.13.
(a) for $t<2600$ (b)for $t<4400$ (c) for $t<20000$.
In (a), color corresponds to the cell's character defined from
the average chemical pattern.  After differentiation,
the activity of a cell is in the order of green, blue, and red while
initial red cells correspond to undifferentiated ones.
The ``green" cell has $\overline{x^{(1)}_i}>.12$, while
the blue cell has $\overline{x^{(2)}_i}<.11$ in Fig. 18b).

\vspace{.1in}
Fig.21

Average chemical concentrations of $x^{(m)}_i$.  
The average is taken over the time steps from the latest cell division
(or since its birth, when it has not experienced the division yet).
The network and parameters are same as for Fig.6.
(a) $t=48$, when the cell number $N=48$, starting from 4 determined cells
and 4 undifferentiated cells
(b) $t=30 $ and $N=32$, starting from 7 determined cells of one type, and
one determined cell of another type.
(c) $t=40$ and $N=42$, starting from 20 determined cells of one type.

Fig.22   

Overlaid time series of $x^{(m)} (t)$ of a
single cell in medium. 
The network is given by (randomly chosen) 4 autocatalytic connections from 64 
chemicals.  The dynamics of
4 chemicals from (a) is given in (b).
The parameters are set as
$p=10.0$, $e_0=e_1=1$, $\overline{X_0}=5$, $\gamma=0.2$, $x_M=10.0$
$D_{out}=f=0.005$, and $V=1000$, while division and death processes are not
included.

\vspace{.1in}

Fig.23  Schematic representation of the evolutionary process of
metabolic networks.  The network (b) is added to (a).
Note that the part (b) is autocatalytic as a set.

\end{document}